\documentclass[epj]{svjour}
\usepackage{graphics}
\begin{document}
\title{Pion-nucleon charge-exchange amplitudes above 2 GeV}
\author{
  F.  Huang\inst{1},
  A.  Sibirtsev\inst{2},
  S.  Krewald\inst{1},
  C.  Hanhart\inst{1},
  J.  Haidenbauer\inst{1},
 \and U.-G. Mei{\ss}ner\inst{1,2}
}                     % Do not remove
%%
%\offprints{S. Krewald}          % Insert a name or remove this line
%
\institute{Institut f\"ur Kernphysik and J\"ulich Center for Hadron Physics, 
Forschungszentrum J\"ulich, D-52425 J\"ulich, Germany \and
Helmholtz-Institut f\"ur Strahlen- und Kernphysik (Theorie) and
Bethe Center for Theoretical Physics, Universit\"at Bonn,
Nu{\ss}allee 14-16, D-53115 Bonn, Germany}
\date{Received: date / Revised version: date}
\abstract{ 
The amplitudes for the pion-nucleon charge exchange reaction of the 
Karlsruhe-Helsinki and the George-Washington-University
partial wave analyses are compared with those of
a Regge-cut model with the aim to explore the possibility to 
provide high energy constraints for
theoretical baryon resonance analyses in the energy region above 2 GeV.
\PACS{
      {13.75.-n}{Hadron-induced low- and intermediate energy reactions}   \and
      {14.20.Gk}{Baryon resonances with S=0}   \and
      {11.55.Jy}{Regge formalism}
     } % end of PACS codes
} %end of abstract
\authorrunning{F. Huang {\it et al.}}
\maketitle

\section{Introduction}
\label{intro}

Presently there is intense experimental activity to study baryon resonances 
in the energy range up to 2.4 GeV
\cite{:2007bk,Aznauryan:2008pe,:2008fz,Sumihama:2007qa}.
Nearly all established resonances summarized in the Review of Particle 
Physics \cite{Yao06} were obtained in the time period before
1980 from partial wave analyses of pion-nucleon reactions by the
Karlsruhe-Helsinki (KH80) \cite{Koch80,Hoehler83} and the Carnegie-Mellon-Berkeley 
(CMB) \cite{Cutkosky:1979fy} collaborations.
Arndt and his collaborators at the George Washington University (GWU)
have published a series of more recent analyses \cite{Arndt04,Arndt06}
including the enlarged data basis provided by LAMPF, TRIUMF, and PSI which 
cast doubt on some of the resonances found in previous analyses.
New data on spin rotation parameters obtained by the ITEP-PNPI
collaborations showed some limitations of the KH80 and
the CMB analyses \cite{alekseev01,alekseev07,Yao06a}.

The experimental data on $\pi \pi N$, $\eta N$, and
  $K\Lambda$ final states require unitary multichannel
resonance parameterizations \cite{Manley:1992yb,Batinic95,Batinic98,Vrana:1999nt}. 
The physical background due to non-resonant processes employed in those
calculations
may show a non-trivial energy dependence which is reflected in the
extracted resonance parameters and calls for further theoretical developments
 \cite{Vrana:1999nt}.
The K-matrix approach to coupled-channel reactions combines  
resonances and non-resonant processes, using effective Lagrangians.
Resonance parameters were extracted using the K-matrix method 
in the energy range up to $\sqrt{s}=2$ GeV \cite{Penner:2002ma}.
The K-matrix approach relies on one major approximation, the omission of the 
real part of the intermediate meson-nucleon propagator. This may affect the 
details of the non-resonant background.
Moreover, it impedes the possibility to generate poles in the S-matrix 
dynamically -- a feature which is strongly emphasized in recent works 
that apply concepts from effective field theories to 
meson-meson and meson-baryon
scattering \cite{Weise,Meissner:1999vr,Oset:2007ex,Kolomeitsev:2003kt}.
Several groups have developped models for
meson-nucleon dynamics based on effective Lagrangians going beyond the K-matrix
approximation and studied pion-nucleon scattering for energies up to around 
2 GeV
\cite{Krehl:1999km,Gasparyan:2003fp,Chen07,JuliaDiaz:2007kz,JuliaDiaz:2007fa}.

An extension of meson-nucleon coupled channel approaches to energies above 
2 GeV is a rather challenging task. In principle, partial 
wave amplitudes would provide the most helpful tool for such an extension.
There are limitations at large energies, however.
In the energy range from 2 GeV to 3.5 GeV, the number of partial waves
employed in the KH80 analysis rises,
as partial waves up to angular momentum $j=37/2$ are necessary.
Naturally, it is difficult to determine such a large number of 
parameters from the available data. 
For very high energies the angular distributions of two-body reactions
show well-known regularities: the differential cross sections 
are dominated by forward scattering. 
Here an economic description of the $t$-dependence of the cross sections 
in forward direction as well as of their energy dependence is given by 
Regge phenomenology \cite{Donnachie:1992ny}, using Regge 
trajectories as basic degrees of freedom. Therefore, the question arises
in how far the amplitudes deduced from such approaches can serve 
as a guideline for the envisaged extension to higher energies. 

In the present paper we survey the available information on the $\pi N$
scattering amplitudes from 2 GeV up to energies where the reaction can be
quantitatively described within Regge phenomenology. Specifically, we consider
the amplitudes that result from the partial wave analyses of the GWU
group (which reaches up to 2.5 GeV) and the KH80 solution (which covers
energies up to 3.49 GeV) and predictions of Regge models, fitted to 
high energy $\pi N$ data. 
We concentrate on the charge exchange pion-nucleon reaction in this work
because of the following reason: 
For Regge pole trajectories, there is a natural ordering according
to the intercept $\alpha ( t=0 )$ of the trajectory \cite{Irving:1977ea}.
The pion-nucleon charge exchange reaction $ \pi^- p \rightarrow \pi^0 n $
is dominated by the $\rho$-trajectory because the selectivity of
the reaction supresses contributions of the Pomeron which is leading for 
the elastic pion-nucleon scattering. 
In view of this we consider a simple Regge-cut model which consists of a 
rho-pole trajectory and a rho-Pomeron cut only and use vertex functions
parameterized by a single exponential.
In addition, we compare with the results of a more sophisticated
Regge parameterization taken from the literature \cite{BARGER}.

We consider the present work as a preparatory step for extending 
coupled-channel approaches of the $\pi N$ interaction to energies 
above 2 GeV. Indeed meson-nucleon coupled channel models and Regge models can 
be considered as two effective theories made for different energy scales  
which employ the degrees of freedom most economical for the energy range considered.
In the case of meson-meson scattering, Regge constraints have been fruitfully
used in the classical and recent Roy equation studies, see e.g. 
\cite{Basdevant:1973ru,Ananthanarayan:2000ht,DescotesGenon:2001tn,Pelaez:2004vs,pelaez:2005}.
It will be interesting to see whether the same can be achieved also in the
context of coupled-channel approaches and the $\pi N$ system.
 
The paper is organized as follows.
In section 2, we summarize the  parameterization of the employed Regge-cut model.
The angular distributions and polarizations obtained are shown in section 3
for some selected energies, followed by a discussion of the energy and momentum
dependence of the differential cross sections. The resulting amplitudes are 
presented and discussed in section 4. Conclusions are drawn in section 5.

\section{Formalism }
 \label{sec:1}

The helicity spin non-flip amplitude ${\cal M}^{++}_\rho$ and spin-flip 
amplitude ${\cal M}^{+-}_\rho$ due to the $\rho$-pole exchange
contribution to the $\pi^- p \to \pi^0 n$ reaction are parametrized
as
\begin{eqnarray}
{\cal M}^{++}_\rho &=& \beta^{++}_\rho \,
{G_\rho(s,t)}\frac{\pi}{\Gamma\!\left[\alpha_\rho(t)\right]}, \\
{\cal M}^{+-}_\rho &=& \sqrt{-t} \,\, \beta^{+-}_\rho \,
{G_\rho(s,t)}\frac{\pi}{\Gamma\!\left[\alpha_\rho(t)\right]},
\end{eqnarray}
where $\beta_\rho$ is the residue function specified later. The
Regge propagator is given by
\begin{eqnarray}
G_\rho(s,t) = \frac{1+\xi_\rho{\rm
exp}\left[-i\pi\alpha_\rho(t)\right]}{{\rm
sin}\!\left[\pi\alpha_\rho(t)\right]}
\left(\frac{q^2}{q_0^2}\right)^{\alpha_\rho(t)},
\label{signature}
\end{eqnarray}
with $\xi_\rho=-1$ being the signature of the $\rho$-trajectory.
The $\rho$ trajectory $\alpha_\rho(t)$ is taken as
\begin{eqnarray}
\alpha_\rho(t) = 1 - \alpha_\rho' m_\rho^2 + \alpha_\rho' t .
\end{eqnarray}
The slope parameter $\alpha_\rho'$ is determined by a fit to the data.
Furthermore, $q$ in Eq.~(\ref{signature}) is the pion momentum in the center-of-mass 
system and $q_0=1$ GeV/c serves as a scale. The factor $\frac{\pi}{\Gamma}$ cancels 
the poles of the Regge propagator (\ref{signature}) in the scattering region. 

\begin{table}[b]
\begin{center}
\caption{Parameters of the $\rho$-pole and 
$\rho$-cut amplitudes.
}
\label{tab:para-1}
\begin{tabular}{|c|c|c|}
\hline%\noalign{\smallskip}
Parameter & $\rho$ & $\rho$-cut \\ 
%\noalign{\smallskip}
\hline%\noalign{\smallskip}
$\beta_0^{++}$ & -23.8$\pm$0.3 & 0.5$\pm$0.3 \\
$b^{++}$ & 2.5$\pm$0.2 & 0.6$\pm$0.4 \\
$\beta_0^{+-}$ & 151.3$\pm$1.1 & -113.7$\pm$2.2 \\
$b^{+-}$ & 1.7$\pm$0.1 & 4.3$\pm$0.3\\
\hline%\noalign{\smallskip}
\multicolumn{3}{|c|}{ $\alpha^\prime_\rho{=}0.81{\pm}0.01$} \\
%\noalign{\smallskip}
\hline%\noalign{\smallskip}
\end{tabular}
\end{center}
\end{table}

The helicity spin non-flip amplitude ${\cal M}^{++}_c$ and spin-flip
amplitude ${\cal M}^{+-}_c$ due to the $\rho$-cut exchange, which
represents the initial and final state interactions, are
parametrized as 
\begin{eqnarray}
{\cal M}^{++}_c &=& \beta^{++}_c \,
{G_c(t,s)}\frac{\pi}{\Gamma\!\left[\alpha_c(t)\right]}\,{\rm ln}^{-1}(s/s_0), \\
{\cal M}^{+-}_c &=& \sqrt{-t} \,\, \beta^{+-}_c \,
{G_c(t,s)}\frac{\pi}{\Gamma\!\left[\alpha_c(t)\right]}\,{\rm ln}^{-1}(s/s_0),
\end{eqnarray}
where the prescription of $G_c(t,s)$ is similar to that of
$G_\rho(t,s)$ except that $\alpha_\rho(t)$ is replaced by
$\alpha_c(t)$. Here $\alpha_c(t)$ is the $\rho$-cut trajectory taken
as
\begin{eqnarray}
\alpha_c(t) = 1 - \alpha_\rho' m_\rho^2 + \frac{\alpha_\rho'
\alpha_P'}{\alpha_\rho' + \alpha_P'}t,
\end{eqnarray}
with $\alpha_P'=0.1$ GeV$^{-2}$ being the slope of the Pomeron
trajectory, which is well defined from the analysis of elastic
scattering data. A scale $s_0 = 1$ GeV has been chosen. 

The residue functions for all amplitudes are parameterized in a
similar way
\begin{eqnarray}
\beta(t)=\beta_0\,{\rm exp}(bt),
\label{vertex}
\end{eqnarray}
where the coupling constant $\beta_0$ and the slope $b$ in the
exponential formfactor are determined by a fit to the data.

The total helicity spin non-flip amplitude ${\cal M}^{++}$ and
spin-flip amplitude ${\cal M}^{+-}$ are given by the sum of 
the above amplitudes, i.e. 
\begin{eqnarray}
{\cal M}^{++} &=& {\cal M}^{++}_\rho + {\cal M}^{++}_c, \\
{\cal M}^{+-} &=& {\cal M}^{+-}_\rho + {\cal M}^{+-}_c.
\end{eqnarray}
The differential cross section is given by
\begin{eqnarray}
\frac{d\sigma}{dt}=\frac{|{\cal M}^{++}|^2+|{\cal
M}^{+-}|^2}{s\,q^2} \ ,
\end{eqnarray}
and the polarization by 
\begin{eqnarray}
P=\frac{2\,{\rm Im}\!\left[{\cal M}^{++}{{\cal
M}^{+-}}^*\right]}{{|{\cal M}^{++}|}^2 + {|{\cal M}^{+-}|}^2} \ .
\end{eqnarray}
The difference of the $\pi^-p$ and $\pi^+p$ total cross sections is  
\begin{eqnarray}
\Delta\sigma\equiv\sigma_{\pi^-p}-\sigma_{\pi^+p}=-\frac{4\sqrt{2\pi}}{q\sqrt{s}}{\rm
Im}{\cal M}^{++}(t=0).
\label{difference}
\end{eqnarray} 
The parameters are listed in Table~\ref{tab:para-1}. Those parameters have
been determined by fitting the data on $\pi^-p\to \pi^0n$ differential cross
sections and polarization for pion beam momenta above 4 GeV/c 
($\sqrt{s} \ge 3$ GeV) and for four-momentum transfer squared 
$|t|\le 2$ GeV$^2$ \cite{Sibirtsev08}.

\section{Cross sections and polarization}

\begin{figure}
\begin{center}
\resizebox{0.45\textwidth}{!}{
  \includegraphics{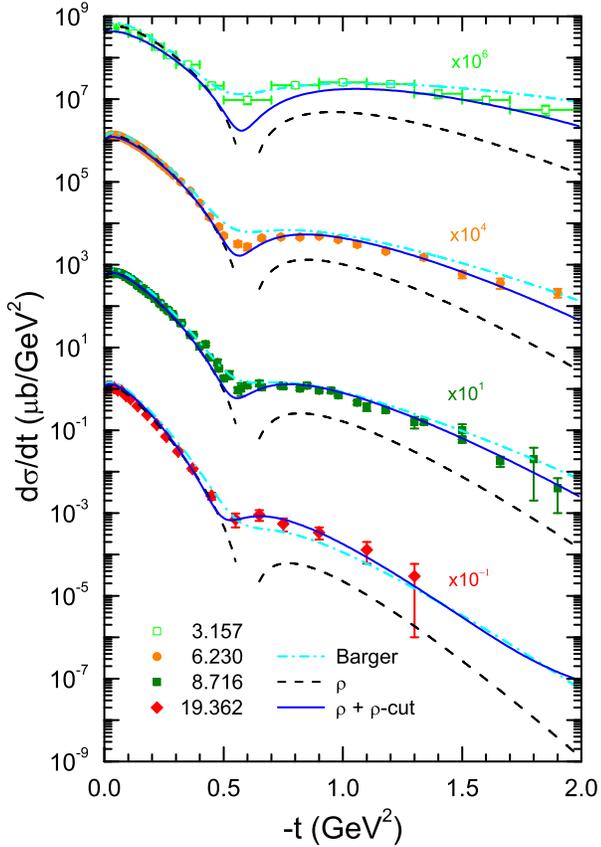}
} 

\caption{
Differential cross sections for the reaction 
$\pi^- p \to \pi^0 n$ as a function of the 
four-momentum transfer squared for different collision
energies indicated in the legend. The solid lines represent the
results of our Regge-cut model,
those based on a pure rho-pole fit are given by the dashed line.
Results for the model of Ref.~\cite{BARGER} are shown as dash-dotted 
lines. The data are taken from Refs.
\cite{Apel77,Apokin82,Barnes76,Bolotov74,Guisan71,Sonderegger66,Wahlig68}.
}
\label{fig:1}
\end{center}
\end{figure}

\begin{figure}
\begin{center}
\resizebox{0.45\textwidth}{!}{
\includegraphics{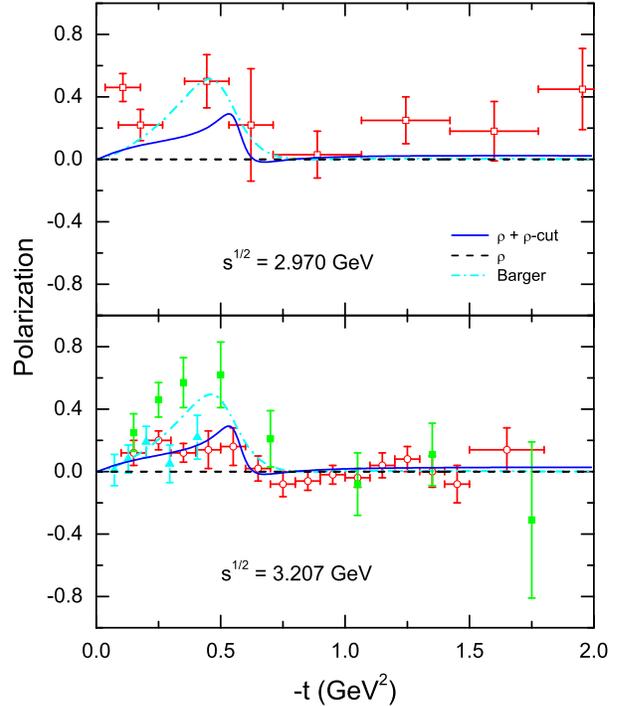}
} 
\caption{The polarization in the reaction $\pi^- p \to \pi^0
n$ as a function of four-momentum transfer
squared for different collision energies indicated
in the legends. Same description of curves as in Fig.~\ref{fig:1}.
The data are taken from Refs. \cite{Minowa87} (open squares),
\cite{Hill73} (circles), \cite{Drobnis68} (triangles), and
\cite{Giacomelli73} (filled squares). 
}
\label{fig:3}
\end{center}
\end{figure}

There are three well-known predictions for the pion-nucleon charge
exchange reaction based on the existence of the rho-trajectory: the
energy dependence of the total cross section, the existence of a
diffractive minimum in the differential cross section in the vicinity 
of $t=-0.6$ GeV$^2$, and the relative phase of the real and imaginary 
parts of the Regge amplitudes which is fixed by 
the signature factor 
$\xi (t) = (1+\xi_\rho{\rm exp}\left[-i\pi\alpha_\rho(t)\right])$ 
(Eq. (\ref{signature})) \cite{Boreskov:1972hv}. 
The Regge limit is supposed to be valid for energies much larger than
the momentum transfers, so that for a given energy, the Regge parameterization
is expected to deteriorate with increasing magnitude of the momentum transfer.
Regge phenomenology modifies the $t$-dependence of the amplitudes determined 
from the Regge trajectory by purely phenomenological vertex functions, 
which allows to optimize the fit to observables. 
For our Regge-cut model we have chosen a rather simple form of the vertex
function, namely an exponential formfactor, cf. Eq.~(\ref{vertex}).
But
in the past, Regge phenomenology has used unconventional parameterizations
of the vertex functions in order to obtain an even faster decrease of the
helicity non-flip amplitude than the one produced by the rho-pole trajectory.
Specifically, functions with a zero crossing at $-t\approx 0.2$ GeV$^2$ 
were employed in order to 
account for the crossover phenomenon. (We will come back to this issue
in the next section when we discuss the amplitudes.) 
One of the models where this has been done is the parameterization of 
Barger and Phillips \cite{BARGER} and, therefore, we will display their
results here for comparison.

The differential cross sections and polarizations obtained are shown in Figs.
\ref{fig:1} and \ref{fig:3} for a few selected energies.
A more systematic
comparison with data will be presented elsewhere \cite{Sibirtsev08}.
Both our Regge-cut model and and the fit of Ref.~\cite{BARGER} reproduce the
data reasonably well. The major difference between the two fits occurs
in the vicinity of the minima close to  $-t=0.6$ GeV$^2$: the Barger-Phillips fit
overestimates the data near the minimum, while the Regge-cut model tends
to lie below the data here. 
In the Regge-cut model, the contribution of the rho-pole dominates 
both the helicity non-flip and the helicity flip amplitudes.
The rho cut amplitudes are a correction which are mainly required to
fill the dip near $-t=0.6$ GeV$^2$. One observes a maximum
of the angular distribution near $-t = 0.03 $ GeV$^2$ which allows
to disentangle the helicity flip and non-flip amplitudes, assuming
the validity of the Regge approach.

For illustration purposes we consider here also results based on
the Regge-pole contribution alone.
The dashed lines in Fig.~\ref{fig:1} have been
obtained using $\beta_0^{++} = -24.62$, $\beta_0^{+-} = 143.29$,
$b^{++} = b^{+-} =2.33$, and  $\alpha_\rho{=}0.832$.
The strength parameters of the flip and non-flip amplitudes for the
pure Regge-pole fit differ by about ten percent from the ones shown in 
Table~\ref{tab:para-1}.
The angular distributions for small values of $-t$ are well reproduced by
the pure Regge-pole model. 
But the description deteriorates rapidly from about $-t=0.4$ GeV$^2$ 
onwards because the pole trajectory enforces a vanishing amplitude at 
$-t=0.6$ GeV$^2$. Still, the pure Regge-pole model produces a second
maximum of the angular distribution, though it underestimates the
magnitude of the corresponding cross section significantly. 

The maximum of the polarization which occurs close to $-t=0.5$~GeV$^2$, c.f. Fig.~\ref{fig:3}, 
is correlated with the first minimum of the angular
distributions, as expected from Eq. (12). Our Regge-cut model predicts a 
smaller polarization than the Barger-Phillips parameterization, but the large 
experimental uncertainties reflected in the various data sets do not allow to 
discriminate between the fits. As is well-known, the pure Regge-pole model predicts 
zero polarization. 

\begin{figure}
\begin{center}
\resizebox{0.45\textwidth}{!}{
  \includegraphics{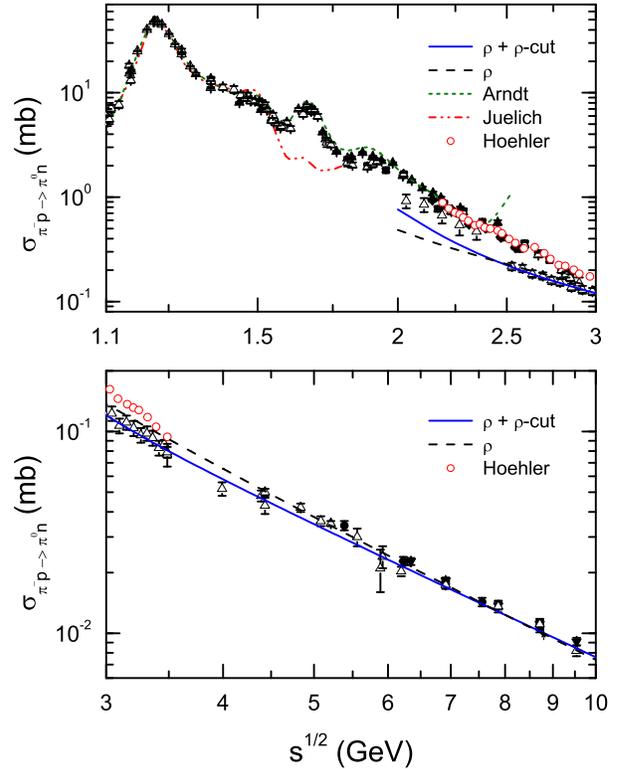}
} \caption{ \label{fig:4} The $\pi^- p \to \pi^0 n$ total cross
sections as a function of the collision energy $\sqrt{s}$. 
The solid and dashed lines are the results of our Regge-cut model
and the pure Regge-pole model, respectively. 
The short-dashed line represents the results of the partial wave
analysis from the GWU group \cite{Arndt06} while the open circles
are those for the KH80 solution \cite{Hoehler83}. 
The dash-dotted line indicates the predictions from the J\"ulich 
meson-exchange model \cite{Gasparyan:2003fp}. 
The data are taken from Refs.
\cite{Apel77,Barnes76,Bolotov74,Giacomelli73,Crouch80,Brown76,Suzuki87}.}
\end{center}
\end{figure}

Fig.~\ref{fig:4} shows the total cross section for the reaction 
$\pi^- p \to \pi^0 n$ as a function of the energy. 
Here we also include results based on two $\pi N$ phase shift analyses, namely
the ones by the GWU \cite{Arndt06} and the Karlsruhe-Helsinki \cite{Hoehler83}
groups. The results for the GWU analysis are those of their current solution
taken from the SAID Program \cite{Said}. With regard to Karlsruhe-Helsinki 
we use the preliminary updated solution KH80 as tabulated in Table 2.2.2.2 of 
Ref.~\cite{Hoehler83}. Though the KH80 solution (and the later KA84 analysis)
is, in principle, available from SAID we used the values from the table
because {\it only there} partial waves amplitudes up to very high angular 
momenta are listed. The SAID Program provides amplitudes only up to
orbital angular momenta of $l = 8$. The KH80 analysis employs partial 
waves with angular momenta up to $l = 11$ around $2.5$ GeV and up to
$l = 18$ at the highest energy of 3.487 GeV \cite{Hoehler83}. 
We found the contributions from those high angular momenta to be 
essential for obtaining converged results for the observables and,
in particular, for reproducing observables as published by H\"ohler 
\cite{Hoehler83}. 
Note that the GWU analysis SP06 which covers energies up to $2.5$ GeV 
\cite{Arndt06} uses partial waves up to $l = 8$.

The Regge-cut model (solid line) is able to reproduce the experimental 
data down to approximately 3 GeV (lower panel of Fig.~\ref{fig:4}). 
Below 3 GeV (upper panel), the 
extrapolation of the Regge-cut model underestimates the
cross section, which suggests the necessity to incorporate
resonances. The energy dependence of the total cross sections can be 
also reproduced by using only the Regge-pole contribution (dashed line).

As far as the partial wave analyses are concerned one can see that
the GWU analysis (short dashed line) starts to deviate from the data 
from around 2.3 GeV onwards, and produces a strongly rising cross section 
not in agreement with the data. This is expected in view of the deteriorating 
$\chi^2$ reported in \cite{Arndt06}. 
The J\"ulich coupled channel meson-exchange model \cite{Gasparyan:2003fp}
includes only a few low-mass resonances and, consequently, does
not agree with the total charge exchange cross sections above $1.5$
GeV, cf. the dash-dotted line. 

Note that there is a conflict between different data sets in the
transition region $2.5 \le \sqrt{s} \le 3$ GeV. The Regge fits tend
to reproduce the smaller values which seem to be more in line with the
high-energy data whereas H\"ohler's analysis agrees with the larger cross
section values in the energy range in question. Those values are compatible 
with the data at lower energies.

\begin{figure}
\begin{center}
\resizebox{0.45\textwidth}{!}{
  \includegraphics{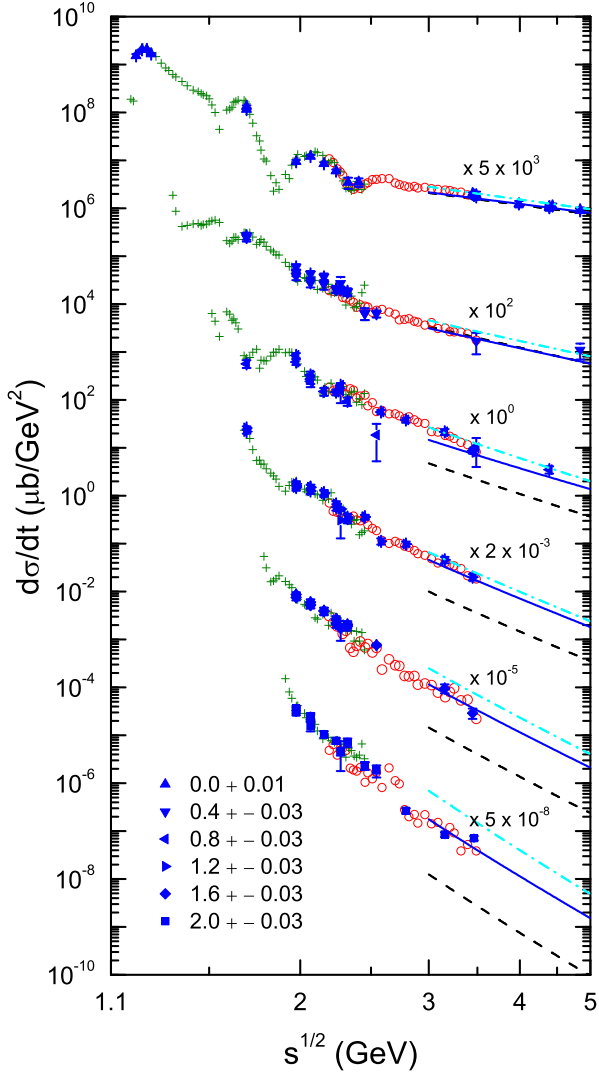}
} \caption{ \label{fig:5} Differential cross sections for the reaction 
$\pi^- p \to \pi^0 n$ at fixed $t$ as a function of the collision energy. 
The $-t$ values considered are 0.0, 0.4, 0.8, 1.2, 1.6, and 2.0 GeV$^2$ from
top to bottom. Same description of curves as in Fig.~\ref{fig:1}. 
The ``+'' symbols represents the results of the partial wave
analysis from the GWU group \cite{Arndt06} while the open circles
are those for the KH80 solution \cite{Hoehler83}. 
The data selected for the various intervals of four-momentum transfers squared, as
denoted in the legend, are taken from Refs.
\cite{Apel77,Apokin82,Barnes76,Bolotov74,Guisan71,Sonderegger66,Wahlig68,Brockett71,Brockett74,Stirling65,Sadler04}.
}
\end{center}
\end{figure}

A systematic view of the
energy dependence of the differential cross sections for several fixed
four-momentum transfers $t$ is provided in Fig.~\ref{fig:5}. Below
$\sqrt{s}\approx 2.5$ GeV, the differential cross sections show
bumps which correspond to known resonances, whereas for 
higher energies the differential cross sections 
decrease smoothly with energy. 
The Regge fits are shown down to 3 GeV.
In the energy range considered, the Regge cuts are an 
important contribution to the cross section.
For small $t$ the KH80 partial wave analysis joins 
the results of the two Regge models.
With increasing $t$ there is a widening gap between the two
Regge models, reflecting the different quality of the fits to the
data at larger $-t$ values. Obviously, the Barger-Phillips fit 
overestimates both the available data and the KH80 analysis for
$-t\approx 1.6-2$ GeV$^2$. On the other hand, our Regge-cut results
are in agreement with the experimental information.

\begin{figure}
\begin{center}
\resizebox{0.5\textwidth}{!}{
  \includegraphics{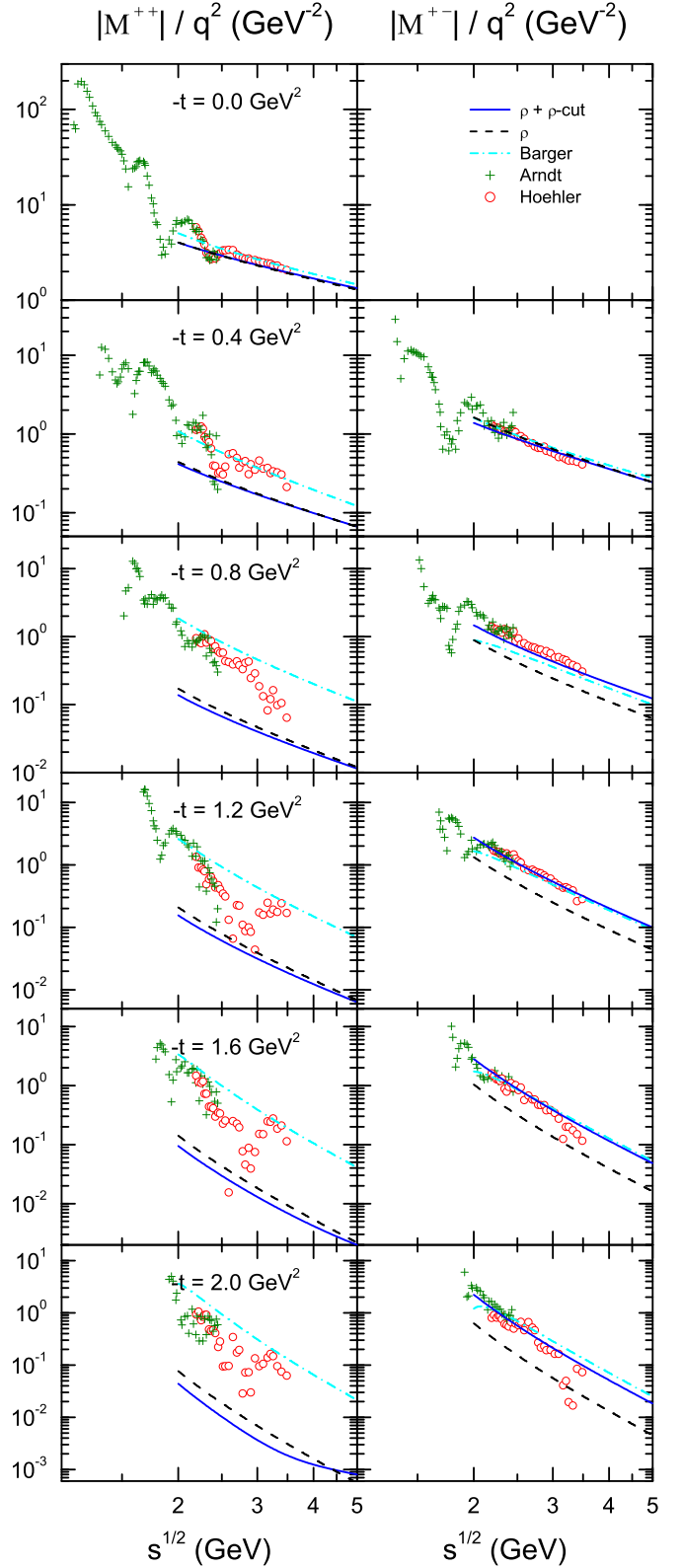}
} \caption{ \label{fig:6} Moduli of the spin non-flip and spin-flip
amplitudes for the reaction $\pi^- p \to \pi^0 n$ 
divided by the center-of-mass momentum squared at fixed $t$ as a
function of the collision energy $\sqrt{s}$. 
Same description of curves as in Fig.~\ref{fig:1}.
The ``+'' symbols and the open circles
denote the results of the partial wave analyses from the
GWU group \cite{Arndt06} and the
Karlsruhe-Helsinki group (KH80) \cite{Hoehler83}, respectively. }
\end{center}
\end{figure}

The helicity flip and non-flip cross sections are observables which
in principle could be measured directly. The differential cross
section for $t=0 $ GeV$^2$ is entirely determined by the helicity 
non-flip transition, which opens the possibility to separate
the two contributions to the cross section
within the Regge model by studying the $t$-dependence of the cross sections.
In Fig.~\ref{fig:6}, we show the energy dependence of the
moduli of the helicity flip and non-flip amplitudes.
The moduli of the helicity flip amplitudes predicted by the fit of 
Ref.~\cite{BARGER} (dash-dotted line) and of our
Regge-cut model (solid line) are close to each other.
The small remaining gap may be taken as a measure for the uncertainty of the
determination of the amplitude. The pure Regge-pole fit produces the correct
energy dependence, but is unable to provide the magnitude required by the 
data for $-t$ larger than 0.4 GeV$^2$, see the dashed line. 
The KH80 analysis joins the Regge fits smoothly, except for some fluctuations
at the largest $|t|$ value considered.

For the helicity non-flip amplitude, the moduli match at
$t= 0$ GeV$^2$, as expected, but at larger four-momentum transfers, considerable 
differences occur. The Barger-Phi{\-}llips fit agrees roughly with
the KH80 analysis for $\sqrt{s} \ge 2.5$ GeV up to $-t\approx 0.8$ GeV$^2$, 
whereas the Regge-cut model produces an amplitude which is much smaller 
than the KH80 result. The helicity non-flip amplitudes of the KH80
analysis
show large fluctuations above 2.5 GeV for all values $-t \ge 0.8 $ GeV$^2$.
One should note that the helicity non-flip amplitude is fairly small above
$\sqrt{s} \approx 2.5$ GeV, which makes its determination difficult.
Direct measurements of polarization and spin rotation parameters 
for forward angles would help.
Recent ITEP experiments have studied other kinematical
regimes \cite{alekseev07}. Pion beams will be also available at J-PARC. 
The HADES collaboration at FAIR/GSI \cite{HADES} prepares a secondary 
pion beam, which would have to be supplemented by polarized targets. 

\begin{figure}
\begin{center}
\resizebox{0.45\textwidth}{!}{
  \includegraphics{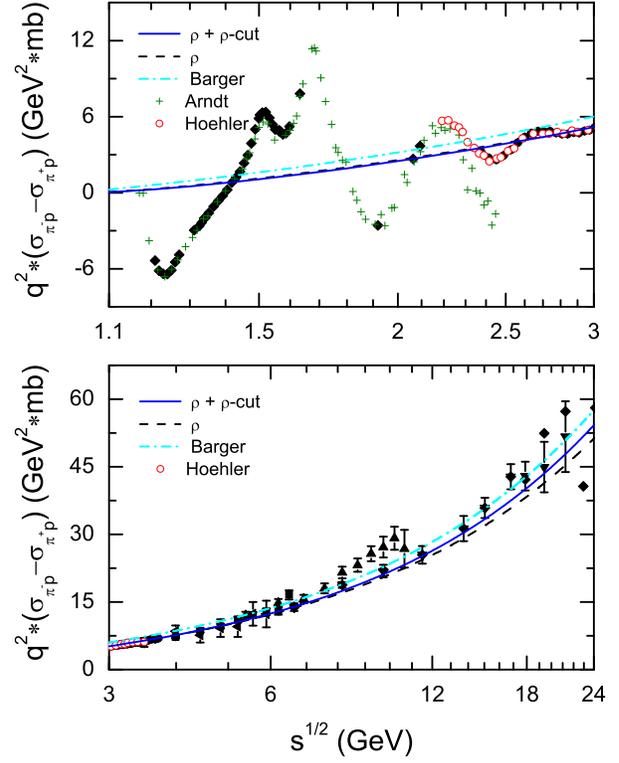}
} \caption{ \label{fig:7} Total cross-section difference for
$\pi^-p$ and $\pi^+p$ scattering multiplied by the three momentum
transfer squared as a function of the collision
energy $\sqrt{s}$. Same description of curves as in Fig.~\ref{fig:1}.
The ``+'' symbols and open circles denote the results
of the partial wave analyses from the GWU group 
\cite{Arndt06} and the Karlsruhe-Helsinki group (KH80)
\cite{Hoehler83}, respectively. Data are taken from Refs.
\cite{Yao06,Carroll74,Carroll76,Citron66,Denisov73,Foley67,Mannelli65}.}
\end{center}
\end{figure}

The total cross-section difference for $\pi^-p$ and $\pi^+p$
scattering is proportional to the imaginary part of the spin
non-flip amplitude for vanishing momentum transfer (see Eq.(13)) and hence can
serve as a test whether the amplitude obtained is acceptable. 
In Fig.~\ref{fig:7}, we show the difference of the cross sections,
multiplied by $q^2$ to facilitate a comparison on a linear scale.
The Regge-cut model (solid line) reproduces the experimental data in the
energy interval ranging from 3 GeV to 24 GeV and appears to average the 
Arndt and H\"ohler analyses below 3 GeV. The partial wave analyses reflect 
the presence of resonances, of course. 
The fact that the Regge amplitude represents quasi an average
of the physical amplitude over the resonance region 
is a manifestation of {\it duality} in strong interaction physics, a 
notion extensively discussed and explored 
in the 1970s (see, e.g., Ref. \cite{Fukugita} for a review), and
recently revived in the context of interrelating quark- and hadronic 
degrees of freedom \cite{Wally}. 

\begin{figure}
\begin{center}
\resizebox{0.5\textwidth}{!}{
  \includegraphics{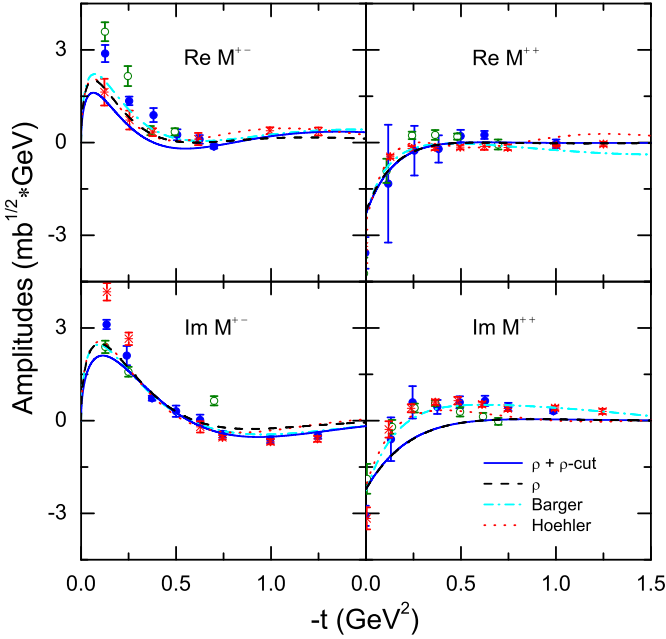}
} 
\caption{ \label{fig:8} Spin-flip and spin non-flip amplitudes 
for the reaction $\pi^- p \to \pi^0 n$ reaction at the collision 
energy $\sqrt{s}=3.487$ GeV as a function of the four-momentum
transfer squared. Same description of curves as in Fig.~\ref{fig:1}.
The dotted lines are the results of the KH80 analysis \cite{Hoehler83}.
The symbols represent results from
different amplitude analyses \cite{Halzen71,Kelly72,Johnson73}.
}
\end{center}
\end{figure}

\begin{figure}
\begin{center}
\resizebox{0.5\textwidth}{!}{
  \includegraphics{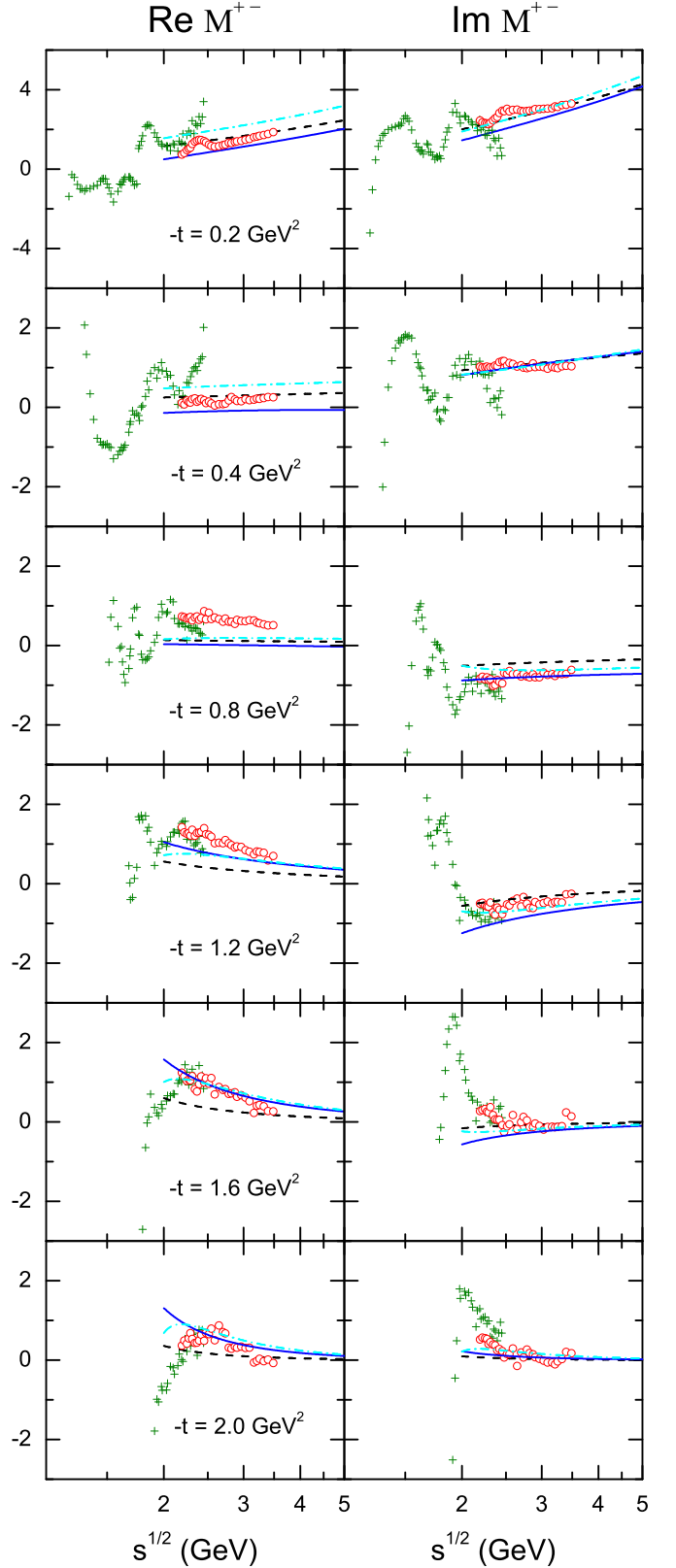}
} \caption{ \label{fig:10} Helicity flip amplitudes for the
reaction $\pi^-p\to\pi^0n$ at fixed $t$ as a
function of the collision energy $\sqrt{s}$. Same description 
of curves as in Fig.~\ref{fig:1}.
The ``+'' symbols and the open
circles denote the results of the partial wave analyses from the
GWU group \cite{Arndt06} and the
Karlsruhe-Helsinki group (KH80) \cite{Hoehler83}, respectively. }
\end{center}
\end{figure}

\section{Amplitudes}

After the measurement of spin rotation parameters for $\pi^+p$ and $\pi^-p$ 
scattering at a beam momentum of 6 GeV/c, corresponding to $\sqrt{s} = 3.49$ GeV, 
several $\pi N$ amplitude analyses were performed at this particular energy 
\cite{Halzen71,Kelly72,Johnson73}.
Since data on $\pi^+p$ and $\pi^-p$ scattering are much more abundant 
(see, e.g., \cite{Arndt06}) and, in general, also more accurate than for 
the charge-exchange channel
the amplitudes of the partial wave analyses are mainly determined by the 
former two reactions. 
On the other hand, in the Regge-model fits considered here only
$\pi N$ charge-exchange data at higher energies have been used as input.
It is therefore interesting to compare systematically the amplitudes derived 
by different methods.

In Fig.~\ref{fig:8}, we contrast the amplitudes
obtained from the present model for the momentum 
$p_{\rm lab} = 6$ GeV$/c$ with three different amplitude
analyses \cite{Halzen71,Kelly72,Johnson73} and the
amplitudes generated from the KH80 analysis. 
The Regge-cute model and the pure rho-pole model produce
very similar amplitudes. 
Both real and imaginary parts of the helicity flip amplitudes are
in reasonable agreement with the amplitude analyses. Note, however, 
that empirically the height of the maximum of the helicity flip amplitude 
near $ -t= 0.2 $ GeV$^2$ is not uniquely determined. 
The magnitude of the helicity non-flip amplitude for vanishing momentum 
transfer is fixed by the total cross section data. But the slope of the
helicity non-flip amplitude near $t=0$ differs from the one obtained in
the Regge-cut model. Specifically, the zero that appears in ${\rm Im} \, M^{++}$
at $t\approx -0.2$ GeV$^2$, which is commonly connected to the 
so-called crossover zero in the $\pi^{\pm}p \to \pi^{\pm}p$ differential
cross sections \cite{Irving:1977ea,Leader73,Goulianos83}, 
is not reproduced by the pure Regge-pole fit but
also not by our Regge-cut model. On the other hand, the result based on the 
H\"ohler analysis (dotted line) clearly exhibits this feature and is also
in good overall agreement with those amplitude analyses.

There have been various suggestions to modify the Regge phenomenology
in order to accomodate a vanishing imaginary
helicity non-flip amplitude for $-t \approx 0.2$ GeV$^2$ \cite{Irving:1977ea}. 
In Ref.~\cite{BARGER}, the vertex function has been
modified to generate a zero at $-t\approx 0.2$ GeV$^2$ in addition to the one produced
by the rho-pole trajectory. The $t$-dependence of the vertex functions assumed 
in Ref.~\cite{BARGER} is entirely phenomenological and might not be optimal 
for other energies. A comparison with our Regge-cut model which assumes
exponentials as vertex functions may therefore be helpful in estimating
the systematic uncertainties in Regge models for the helicity non-flip
amplitude. One can see in Fig.~\ref{fig:8} that the Regge fit by Barger
and Phillips describes the zero in ${\rm Im} \, M^{++}$ at small $-t$ (dash-dotted line). 
Note that we have changed the signs of their predictions for ${\rm Im} \, M^{+-}$
and ${\rm Re} \, M^{++}$ (here and also below) in order to make them comparable to the 
other results. 
It remains unclear to us whether this is a matter of conventions only or whether
the Barger-Phillips parameterization~\cite{BARGER} produced indeed different
signs for those amplitudes. 

\begin{figure}
\begin{center}
\resizebox{0.5\textwidth}{!}{
  \includegraphics{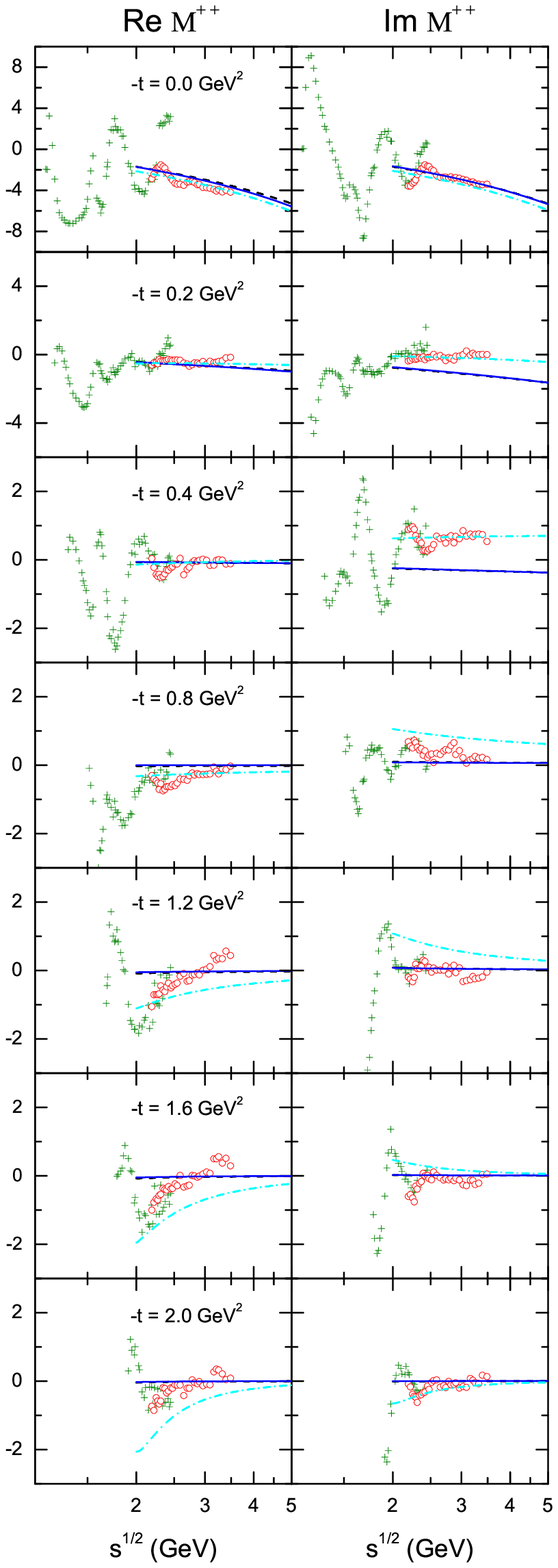}
} \caption{ \label{fig:9} Helicity non-flip amplitudes. See Fig.~\ref{fig:10}
for details.
 }
\end{center}
\end{figure}

A systematic comparison of the energy dependence of the amplitudes
is presented in Figs.~\ref{fig:10} and \ref{fig:9}. Let us first
discuss the real and imaginary parts of the helicity flip amplitudes
which are shown in Fig.~\ref{fig:10}.
For energies from around $\sqrt{s}\approx 2.3 $ GeV onwards the GWU analysis
and the KH80 analysis start to deviate from each other.
Our Regge-cut model and the Barger-Phillips parameterization use quite 
different phenomenological assumptions, yet they produce very similar imaginary 
parts for the helicity flip amplitude for energies above $\sqrt{s} \approx 3$~GeV. 
Also the KH80 result is close to the Regge models in this energy region for 
practically all $t$-values considered.
The real parts of the amplitudes generated by those two Regge fits agree for 
momentum transfer $-t\ge 0.8$ GeV$^2$, but show noticeable differences for 
small $|t|$-values. Moreover, in the latter region neither of them is
in line with the KH80 result. The largest differences in the
real part of the helicity flip amplitude between the KH80 analysis and
the Regge models occurs around $-t\approx 0.8$ GeV$^2$. 
Interestingly, the pure Regge-pol fit produces a helicity flip amplitude
quite close to the one of the KH80 analysis for $-t\le 0.4$ GeV$^2$
(for the real as well as imaginary part), 
cf. the dashed line in Fig.~\ref{fig:10}.

The amplitudes for the helicity non-flip amplitudes are shown in Fig.~\ref{fig:9}.
Also here one observes a deviation of the
GWU amplitudes from the KH80 result from around 2.3 GeV onwards.
For $t=0$ GeV$^2$, the KH80 analysis and the two Regge models are in reasonable
agreement for energies above $ \sqrt{s}\approx 2.5$ GeV. Note that 
${\rm Im} \, M^{++} (t=0)$ is proportional to the difference of the
$\pi^-p$ and $\pi^+p$ cross section (cf. Eq.~(\ref{difference})) and, therefore, 
an empirically accessible quantity. 
For $ -t\approx 0.2$ GeV$^2$, the helicity non-flip amplitude of the 
Barger-Phillips fit vanishes by construction and is here in line with 
the results of the KH80 analysis. 
The Regge-cut model employs an exponential vertex function and obtains an
imaginary amplitude different from the other analyses.
For $-t \ge 0.8$ GeV$^2$, the situation changes: here the Barger-Phillips prediction
for ${\rm Im} \, M^{++}$ is larger than the KH80 analysis for all energies, whereas 
the result of our Regge-cut model approaches zero as suggested by the Karlsruhe 
Helsinki analysis. Suprisingly, the real part of $M^{++}$ of the KH80 analysis 
exhibits a significant energy dependence for fairly large values of $|t|$ 
which is neither reproduced by our Regge-cut model nor by the Barger-Phillips
parameterization. 

\section{Conclusions}

The experimental search for baryon resonances with masses above 2
GeV requires theoretical analyses based on coupled reaction channel
approaches. 
Partial wave amplitudes provide a convenient summary of the experimental 
data, but may suffer from convergence problems at large energies. 
In the high energy limit, two-hadron reactions are diffractive and 
can be described economically and quantitatively within Regge
phenomenology. The corresponding amplitudes offer the possibility to 
obtain constraints for theoretical approaches from the high energy 
region. 
Thus, in lieu of partial waves, amplitudes for forward scattering as
predicted by Reggy models may provide an interface between theory and 
data.

In this paper, we presented a Regge-cut model which reproduces the 
differential cross sections down to energies $\sqrt{s}\approx 3$ GeV 
and for momentum transfer $ -t < 2$ GeV$^2$.
We then compared the resulting amplitudes with those determined in the
Karlsruhe-Helsinki (KH80) partial wave analysis. 
It turned out that 
the magnitudes of the helicity non-flip amplitudes are not well constrained. 
There is strong evidence for a rapid decrease with increasing momentum transfer.
The amplitudes of the Barger-Phillips fit \cite{BARGER}, which
we considered here as example of an alternative Regge parameterization, 
may be taken as an estimate for the upper limit of the helicity non-flip 
amplitude.
The magnitudes of the helicity flip amplitudes derived from the
Regge-cut model join the corresponding quantities obtained in the KH80
partial wave analysis smoothly in the vicinity of $\sqrt{s}=3$~GeV. 
We conclude that the appropriate energy region for matching 
meson-nucleon dynamics to diffractive scattering could be around 
approximately 3 GeV for the helicity flip amplitude of the 
$\pi N$ charge-exchange reaction.

\begin{acknowledgement}
This work is partially supported by the Helmholtz Association
through funds provided to the virtual institute ``Spin and strong
QCD'' (VH-VI-231), by the EU Integrated Infrastructure Initiative
Hadron Physics Project under contract number RII3-CT-2004-506078
and by DFG (SFB/TR 16, ``Subnuclear Structure of Matter'').
F. H. is grateful for the support from the Alexander von Humboldt Foundation.
A.S. acknowledges support by the JLab grant SURA-06-C0452 and the 
COSY FFE grant No. 41760632 (COSY-085).
\end{acknowledgement}


\begin{thebibliography}{}

%\cite{:2007bk}
\bibitem{:2007bk}
  A.~V.~Sarantsev {\it et al.},
  %``New results on the Roper resonance and the $P_{11}$ partial wave,''
  Phys.\ Lett.\  B {\bf 659}, 94 (2008)
  [arXiv:0707.3591 [hep-ph]].
  %%CITATION = PHLTA,B659,94;%%

%\cite{Aznauryan:2008pe}
\bibitem{Aznauryan:2008pe}
  I.~G.~Aznauryan {\it et al.}  [CLAS Collaboration],
  %``Electroexcitation of the Roper resonance for 1.7 < Q2 < 4.5 GeV2 in ep ->
  %enpi+,''
  arXiv:0804.0447 [nucl-ex].
  %%CITATION = ARXIV:0804.0447;%%

%\cite{:2008fz}
\bibitem{:2008fz}
  R.~Nasseripour {\it et al.}  [CLAS Collaboration],
  %``Polarized Structure Function $\sigma_{LT'}$ for $p({\vec e},e'K^+)\Lambda$
  %in the Nucleon Resonance Region,''
  Phys.\ Rev.\  C {\bf 77}, 065208 (2008)
  [arXiv:0801.4711 [nucl-ex]].
  %%CITATION = PHRVA,C77,065208;%%

%\cite{Sumihama:2007qa}
\bibitem{Sumihama:2007qa}
  M.~Sumihama {\it et al.},
  %``Backward-angle photoproduction of $\pi^0$ mesons on the proton at
  %$E_\gamma$ = 1.5--2.4 GeV,''
  Phys.\ Lett.\  B {\bf 657}, 32 (2007)
  [arXiv:0708.1600 [nucl-ex]].
  %%CITATION = PHLTA,B657,32;%%

\bibitem{Yao06}
C. Amsler {\it et al.}, Phys. Lett. B {\bf 667}, 1 (2008). 

\bibitem{Koch80}
R. Koch and E. Pietarinen, Nucl. Phys. A {\bf 336}, 331 (1980). 

\bibitem{Hoehler83}
G. H\"ohler, Landolt-B\"ornstein Series, Group I, Elementary
Particles, Nuclei and Atoms {\bf 9b1}, Springer, Berlin, 1983.

%\cite{Cutkosky:1979fy}
\bibitem{Cutkosky:1979fy}
R.~E.~Cutkosky, C.~P.~Forsyth, R.~E.~Hendrick and R.~L.~Kelly,
%``Pion - Nucleon Partial Wave Amplitudes,''
Phys.\ Rev.\  D {\bf 20}, 2839 (1979).
%%CITATION = PHRVA,D20,2839;%%

\bibitem{Arndt04}
  R.~A.~Arndt, W.~J.~Briscoe, I.~I.~Strakovsky, R.~L.~Workman and M.~M.~Pavan,
  %``Dispersion relation constrained partial wave analysis of pi N elastic  and
  %pi N --> eta N scattering data: The baryon spectrum,''
  Phys.\ Rev.\  C {\bf 69}, 035213 (2004)
  [arXiv:nucl-th/0311089].

\bibitem{Arndt06}
R.A. Arndt, W.J. Briscoe, I.I. Strakovsky and R.L. Workman, Phys.
Rev. C {\bf 74}, 045205 (2006).

\bibitem{Said}
        R.A. Arndt, W.J. Briscoe, R.L. Workman and I.I. Strakovsky, CNS Data
        Analysis Center, {\it http://gwdac.phys.gwu.edu/}

\bibitem{alekseev01}
I.G. Alekseev {\it et al.}, Eur. Phys. J. A {\bf 12}, 117 (2001). 

\bibitem{alekseev07}
  I.~G.~Alekseev {\it et al.},
  %``Asymmetry measurement in the elastic pion-proton scattering at 1.94-GeV/c
  %and 2.07-GeV/c,''
  AIP Conf.\ Proc.\  {\bf 915}, 665 (2007);
  arXiv:0810.1143 [hep-ex].
\bibitem{Yao06a} See the comments of G. H\"ohler and R.L. Workman in 
W.M. Yao {\it et al.}, J. Phys. G {\bf 33}, 1 (2006).

%\cite{Manley:1992yb}
\bibitem{Manley:1992yb}
  D.~M.~Manley and E.~M.~Saleski,
  %``Multichannel Resonance Parametrization Of Pi N Scattering Amplitudes,''
  Phys.\ Rev.\  D {\bf 45}, 4002 (1992).
  %%CITATION = PHRVA,D45,4002;%%

\bibitem{Batinic95}
  M.~Batini\'c, I.~\v Slaus, A.~\v Svarc and B.~M.~K.~Nefkens,
  %``pi N $\to$ eta N and eta N $\to$ eta N partial wave T matrices in a
  %coupled, three channel model,''
  Phys.\ Rev.\  C {\bf 51}, 2310 (1995)
  [Erratum-ibid.\  C {\bf 57}, 1004 (1998)]
  [arXiv:nucl-th/9501011].

\bibitem{Batinic98}
  M.~Batini\'c, I. Dadi\'c, I. \v Slaus, A.~\v Svarc, B.~M.~K.~Nefkens and T.~S.~H.~Lee,
  %``Near threshold eta production in proton proton collisions,''
  Phys.\ Scripta {\bf 58}, 15 (1998).

%\cite{Vrana:1999nt}
\bibitem{Vrana:1999nt}
  T.~P.~Vrana, S.~A.~Dytman and T.~S.~H.~Lee,
  %``Baryon resonance extraction from pi N data using a unitary multichannel
  %model,''
  Phys.\ Rept.\  {\bf 328}, 181 (2000)
  [arXiv:nucl-th/9910012].
  %%CITATION = PRPLC,328,181;%%

%\cite{Penner:2002ma}
\bibitem{Penner:2002ma}
  G.~Penner and U.~Mosel,
  %``Vector meson production and nucleon resonance analysis in a coupled
  %channel approach for energies m(N) < s**(1/2) < 2-GeV. I: Pion induced
  %results and hadronic parameters,''
  Phys.\ Rev.\  C {\bf 66}, 055211 (2002)
  [arXiv:nucl-th/0207066].
  %%CITATION = PHRVA,C66,055211;%%

\bibitem{Weise} 
N.~Kaiser, T.~Waas and W.~Weise,
Nucl.\ Phys.\ A {\bf 612}, 297 (1997).

%\cite{Meissner:1999vr}
\bibitem{Meissner:1999vr}
U.-G.~Mei{\ss}ner and J.~A.~Oller,
Nucl.\ Phys.\ A {\bf 673}, 311 (2000).

%\cite{Oset:2007ex}
\bibitem{Oset:2007ex}
  E.~Oset {\it et al.},
  %``Photo- and Electron-Production of Mesons on Nucleons and Nuclei,''
  Prog.\ Part.\ Nucl.\ Phys.\ {\bf 61}, 260 (2008)
  [arXiv:0711.2967 [nucl-th]].
  %%CITATION = PPNPD,61,260;%%

%\cite{Kolomeitsev:2003kt}
\bibitem{Kolomeitsev:2003kt}
  E.~E.~Kolomeitsev and M.~F.~M.~Lutz,
  %``On baryon resonances and chiral symmetry,''
  Phys.\ Lett.\  B {\bf 585}, 243 (2004)
  [arXiv:nucl-th/0305101].
  %%CITATION = PHLTA,B585,243;%%

%\cite{Krehl:1999km}
\bibitem{Krehl:1999km}
  O.~Krehl, C.~Hanhart, S.~Krewald and J.~Speth,
  %``What is the structure of the Roper resonance?,''
  Phys.\ Rev.\  C {\bf 62}, 025207 (2000)
  [arXiv:nucl-th/9911080].
  %%CITATION = PHRVA,C62,025207;%%

%\cite{Gasparyan:2003fp}
\bibitem{Gasparyan:2003fp}
  A.~M.~Gasparyan, J.~Haidenbauer, C.~Hanhart and J.~Speth,
  %``Pion nucleon scattering in a meson exchange model,''
  Phys.\ Rev.\  C {\bf 68}, 045207 (2003)
  [arXiv:nucl-th/0307072].
  %%CITATION = PHRVA,C68,045207;%%

\bibitem{Chen07}
  G.~Y.~Chen, S.~S.~Kamalov, S.~N.~Yang, D.~Drechsel and L.~Tiator,
  %``Nucleon resonances in \pi N scattering up to energies \sqrt(s) < 2.0 GeV,''
  Phys.\ Rev.\  C {\bf 76}, 035206 (2007)
  [arXiv:nucl-th/0703096].

%\cite{JuliaDiaz:2007kz}
\bibitem{JuliaDiaz:2007kz}
  B.~Julia-Diaz, T.~S.~Lee, A.~Matsuyama and T.~Sato,
  %``Dynamical Coupled-Channel Model of $\pi N$ Scattering in the W $\leq$ 2 GeV
  %Nucleon Resonance Region,''
  Phys.\ Rev.\  C {\bf 76}, 065201 (2007)
  [arXiv:0704.1615 [nucl-th]].
  %%CITATION = PHRVA,C76,065201;%%

%\cite{JuliaDiaz:2007fa}
\bibitem{JuliaDiaz:2007fa}
  B.~Julia-Diaz, T.~S.~Lee, A.~Matsuyama, T.~Sato and L.~C.~Smith,
  %``Dynamical Coupled-Channels Effects on Pion Photoproduction,''
  Phys.\ Rev.\  C {\bf 77}, 045205 (2008)
  [arXiv:0712.2283 [nucl-th]].
  %%CITATION = PHRVA,C77,045205;%%

%\cite{Donnachie:1992ny}
\bibitem{Donnachie:1992ny}
  A.~Donnachie and P.~V.~Landshoff,
  %``Total cross-sections,''
  Phys.\ Lett.\  B {\bf 296}, 227 (1992)
  [arXiv:hep-ph/9209205].
  %%CITATION = PHLTA,B296,227;%%

%\cite{Irving:1977ea}
\bibitem{Irving:1977ea}
  A.~C.~Irving and R.~P.~Worden,
  %``Regge Phenomenology,''
  Phys.\ Rept.\  {\bf 34}, 117 (1977).
  %%CITATION = PRPLC,34,117;%%

%\cite{BARGER}
\bibitem{BARGER}
V. Barger and R.J.N. Phillips, 
Phys. Lett. B {\bf 53}, 195 (1974).

%\cite{Basdevant:1973ru}
\bibitem{Basdevant:1973ru}
  J.~L.~Basdevant, C.~D.~Froggatt and J.~L.~Petersen,
  %``Construction Of Phenomenological Pi Pi Amplitudes,''
  Nucl.\ Phys.\  B {\bf 72} (1974) 413.
  %%CITATION = NUPHA,B72,413;%%

%\cite{Ananthanarayan:2000ht}
\bibitem{Ananthanarayan:2000ht}
  B.~Ananthanarayan, G.~Colangelo, J.~Gasser and H.~Leutwyler,
  %``Roy equation analysis of pi pi scattering,''
  Phys.\ Rept.\  {\bf 353} (2001) 207
  [arXiv:hep-ph/0005297].
  %%CITATION = PRPLC,353,207;%%

%\cite{DescotesGenon:2001tn}
\bibitem{DescotesGenon:2001tn}
  S.~Descotes-Genon, N.~H.~Fuchs, L.~Girlanda and J.~Stern,
  %``Analysis and interpretation of new low-energy pi pi scattering data,''
  Eur.\ Phys.\ J.\  C {\bf 24} (2002) 469
  [arXiv:hep-ph/0112088].
  %%CITATION = EPHJA,C24,469;%%

%\cite{Pelaez:2004vs}
\bibitem{Pelaez:2004vs}
  J.~R.~Pelaez and F.~J.~Yndurain,
  %``The pion pion scattering amplitude,''
  Phys.\ Rev.\  D {\bf 71}, 074016 (2005)
  [arXiv:hep-ph/0411334].
  %%CITATION = PHRVA,D71,074016;%%
\bibitem{pelaez:2005}
 J.~R.~Pelaez and F.~J.~Yndurain,
 Phys.\ Rev.\  D {\bf 69}, 114001 (2004).

\bibitem{Sibirtsev08} A. Sibirtsev {\it et al.}, in preparation. 

%\cite{Boreskov:1972hv}
\bibitem{Boreskov:1972hv}
  K.~G.~Boreskov, A.~M.~Lapidus, S.~T.~Sukhorukov and K.~A.~Ter-Martirosyan,
  %``High energy scattering and charge exchanges in the complex angular momentum
  %theory,''
  Nucl.\ Phys.\  B {\bf 40}, 307 (1972).
  %%CITATION = NUPHA,B40,307;%%

\bibitem{Apel77}
W.D. Apel {\it et al.}, Phys. Lett. B {\bf 72}, 132 (1977).

\bibitem{Apokin82}
V.D. Apokin {\it et al.}, Z. Phys. C {\bf 15}, 293 (1982).

\bibitem{Barnes76}
A.V. Barnes {\it et al.}, Phys. Rev. Lett. {\bf 37}, 76 (1976).

\bibitem{Bolotov74}
V.N. Bolotov {\it et al.}, Nucl. Phys. B {\bf 73}, 365 (1974).

\bibitem{Guisan71}
O. Guisan, P. Bonamy, P.Le Du and L. Paul, Nucl. Phys. B {\bf 32},
681 (1971).

\bibitem{Sonderegger66}
P. Sonderegger {\it et al.}, Phys. Lett. {\bf 20}, 75 (1966).

\bibitem{Wahlig68}
M.A. Wahlig and I. Mannelli, Phys. Rev. {\bf 168}, 1515 (1968).

\bibitem{Minowa87}
M. Minowa {\it et al.}, Nucl. Phys. B {\bf 294}, 979 (1987).

%\bibitem{Bonamy73}
%P. Bonamy {\it et al.}, Nucl. Phys. B {\bf 52}, 392 (1973).

\bibitem{Hill73}
D. Hill {\it et al.}, Phys. Rev. Lett. {\bf 30}, 239 (1973).

\bibitem{Drobnis68}
D.D. Drobnis {\it et al.}, Phys. Rev. Lett. {\bf 20}, 274 (1968).

\bibitem{Giacomelli73}
G. Giacomelli, Landolt-B\"ornstein Series, Group I, Elementary
Particles, Nuclei and Atoms {\bf 7}, Springer, Berlin, 1973.

%\bibitem{Bonamy70}
%P. Bonamy {\it et al.}, Nucl. Phys. B {16}, 335 (1970).

\bibitem{Crouch80}
H.R. Crouch {\it et al.}, Phys. Rev. D {\bf 21}, 3023 (1980).

\bibitem{Brown76}
R.M. Brown {\it et al.}, Nucl. Phys. B {\bf 117}, 12 (1976).

\bibitem{Suzuki87}
Y. Suzuki {\it et al.}, Nucl. Phys. B {\bf 294}, 961 (1987).


\bibitem{Brockett71}
W.S. Brockett {\it et al.}, Phys. Rev. Lett. {\bf 26}, 527 (1971).

\bibitem{Brockett74}
W.S. Brockett {\it et al.}, Phys. Lett. B {\bf 51}, 390 (1974).

\bibitem{Stirling65}
A.V. Stirling {\it et al.}, Phys. Rev. Lett. {\bf 14}, 763 (1965).

\bibitem{Sadler04}
M.E. Sadler {\it et al.}, Phys. Rev. C {\bf 69}, 055206 (2004).

\bibitem{HADES} Hades Collaboration, {\it http://www-hades.gsi.de/}

\bibitem{Carroll74}
A.S. Carroll {\it et al.}, Phys. Rev. Lett. {\bf 33}, 928 (1974).

\bibitem{Carroll76}
A.S. Carroll {\it et al.}, Phys. Lett. B {\bf 61}, 303 (1976).

\bibitem{Citron66}
A. Citron {\it et al.}, Phys. Rev. {\bf 144}, 1101 (1966).

\bibitem{Denisov73}
S.V. Denisov {\it et al.}, Nucl. Phys. B {\bf 65}, 1 (1973).

\bibitem{Foley67}
K.J. Foley {\it et al.}, Phys. Rev. Lett. {\bf 19}, 857 (1967).

\bibitem{Mannelli65}
I. Mannelli, A. Bigi, R. Carrara, M. Wahlig and L. Sodickson, Phys.
Rev. Lett. {\bf 14}, 408 (1965).

\bibitem{Fukugita}
M. Fukugita and K. Igi, Phys. Rept. {\bf 31}, 237 (1977). 

\bibitem{Wally}
  W.~Melnitchouk, R.~Ent and C.~Keppel,
  %``Quark-hadron duality in electron scattering,''
  Phys.\ Rept.\  {\bf 406}, 127 (2005)
  [arXiv:hep-ph/0501217].

\bibitem{Halzen71}
F. Halzen and C. Michael, Phys. Lett. B {\bf 36}, 367 (1971).

\bibitem{Kelly72}
R.L. Kelly, Phys. Lett. B {\bf 39}, 635 (1972).

\bibitem{Johnson73}
P. Johnson, K.E. Lassila, P. Koehler, R. Miller and A. Yokosawa,
Phys. Rev. Lett. {\bf 30}, 242 (1973).

\bibitem{Leader73}
E. Leader and B. Nicolescu,
Phys. Rev. D {\bf 7}, 836 (1973).

\bibitem{Goulianos83}
K. Goulianos, Phys. Rept. {\bf 101}, 169 (1983).

\end{thebibliography}
\end{document}